# Multiple structure and symmetry types in narrow temperature and magnetic field ranges in two-dimensional $Cr_2Ge_2Te_6$ crystal

Tengfei Guo[1, 3, a)], Zongwei Ma[1, a)], Gaoting Lin[2, 3, a)], Xuan Luo[2], Yubin Hou[1], Yuping Sun[1, 2], Zhigao Sheng[1, *], Qingyou Lu[1,5,6*]

[1]*Anhui Province Key Laboratory of Condensed Matter Physics at Extreme Conditions, High Magnetic Field Laboratory, Chinese Academy of Sciences, Hefei, 230031, China*

[2]*Key Laboratory of Materials Physics, Institute of Solid State Physics, Chinese Academy of Sciences, Hefei, 230031, China*

[3]*University of Science and Technology of China, Hefei, 230026, China*

[4]*Institute of Solid State Physics, Chinese Academy of Sciences, Hefei, 230031, China*

[5]*Hefei National Laboratory for Physical Sciences at Microscale, University of Science and Technology of China, Hefei, 230026, China*

[6]*Collaborative Innovation Center of Advanced Microstructure, Najing University, Nanjing 210093, China*

**Abstract:** Multiple structure and symmetry types and their transformations are discovered in bulk two-dimensional (2D) $Cr_2Ge_2Te_6$ crystal within surprisingly narrow temperature range of 2 K and magnetic field range of 0.07 T using a homebuilt magnetic force microscope (MFM). A series of basic domain patterns are extracted from the MFM images. Some of them seem unique to 2D materials as they are not observed in 3D materials, such as self-fitting disks, and fine ladder structure within Y-connected walls. Based on these findings, a phase map is drawn for the magnetic phase structures. The symmetry of these patterns are discussed. The results are not only important in developing new theories but also highly desirable in applications.

## 1 Introduction

Two-dimensional (2D) materials have drawn wide attention and been investigated for its novel properties and potential applications[1,2,3] since the discovery

---
a) These authors contributed equally to this work

of graphene[4]. Among them, 2D magnetic semiconductors are promising materials for spintronics applications[5] because they open up a new world of optoelectronic and nano-spintronic devices with smaller size and higher storage density[6,7,8].

Although spontaneous ferromagnetism is impossible in isotropic 2D Heisenberg spin system according to Mermin-Wagner-Berezinskii theorem, the introduction of anisotropy by interlayer coupling, single-ion anisotropy, or external magnetic field, etc., still sustains long-range ferromagnetic order[9]. For example, in $Cr_2Ge_2Te_6$ of which the intrinsic magnetism has recently been confirmed to survive in the 2D limit, the interlayer coupling and external magnetic field induced anisotropy has an important effect on the determination of long-range magnetic order phase transition temperature $T_C$[10]. Critical behavior around $T_C$ suggests ambiguously the 2D nature of bulk $Cr_2Ge_2Te_6$'s intrinsic ferromagnetism[11] which people believe is dominated by strong but short-range intralayer spin exchange. It does not conflict with the fact that interlayer coupling intensively improves the bulk's Curie temperature[10], since interlayer coupling in some way protects the long-range order from being interrupted by thermal fluctuation. Therefore, bulk $Cr_2Ge_2Te_6$ crystal in the magnetic phase transition region provides a good platform for investigating the combined effect of thermal fluctuation, interlayer coupling, and external magnetic field on 2D long-range magnetic order.

Magnetic domains are the embodiment of long-range magnetic ordering in real-space. The formation and evolution of the domain structure are greatly affected by thermal fluctuations in quasi-2D spin systems[12]. The 2D ferromagnetic domains are expected to present changeful pattern types and symmetries in the phase transition region around $T_C$ under daedal effects of in-plane exchange interaction, interlayer coupling, thermal fluctuations and external magnetic field. However, this real-space domain structure cannot be sufficiently resolved by macroscopic SQUID measurements or even microscopic optical measurements with only sub-micrometer resolution, though some indication of new phenomenon in the microscopic 2D magnetism has been observed very recently[13]. It still calls for high spatially resolved imaging of the real-space domain structures and revealing their symmetry features to help establish more reasonably accurate theory with respect to the 2D long-range magnetic ordering.

In this work, using our homebuilt variable temperature and field magnetic force microscope (MFM) with high sensitivity and nanoscale resolution, we have achieved the success in visualizing the fine details of the magnetic structure in a bulk $Cr_2Ge_2Te_6$ sample and how they react to the variations of magnetic field and temperature. It turns out that there exist plenty of different types of magnetic structures and symmetries in a very narrow temperature range of 2 K and a very narrow magnetic field range of 0.07 T, which may be particularly important in practical applications like temperature or magnetic field sensors. Some of the structures could be unique to 2D materials, for example self-fitting disks and fine ladder structure within the Y-connected walls, etc., because they have not been previously reported in 3D materials to the best of our knowledge. A complicated phase diagram is drawn on the basis of extraction and classification of all the observed domain patterns. These rich discoveries in a relatively simple system will surely provide a solid foundation for future theoretical studies, which is crucial because even at the present, little is known about the microscopic domain structure types and symmetries in real intrinsic 2D ferromagnets, especially the evolution behavior of the domains under external magnetic field and in the magnetic phase transition region around Curie temperature.

**2 Experiments**

The layered $Cr_2Ge_2Te_6$ crystal sample is grown by the self-flux method and the details can be found in the reported paper[14]. The size of the studied crystal with a shining surface is about 3×3×0.5 $mm^3$. The single crystals are air-stable and cleaved before the measurements. Its optical photograph and crystal structure are present in Fig. 1 (a) and (b), respectively. The bulk magnetization behavior was characterized by SQUID measurements of *M-T* and *M-H* curves, which presents a Curie temperature around 65 K and a smooth variation of macroscopic magnetization versus temperature and magnetic field applied along the *c* axis of the crystal, as shown in Fig. 1 (c) and

(d). The macroscopic magnetization data does not give any useful information about the dependence of real-space resolved magnetic domain structure type or symmetry evolution on varying temperature or magnetic field. So we turn to direct MFM observation of the domain patterns under different conditions.

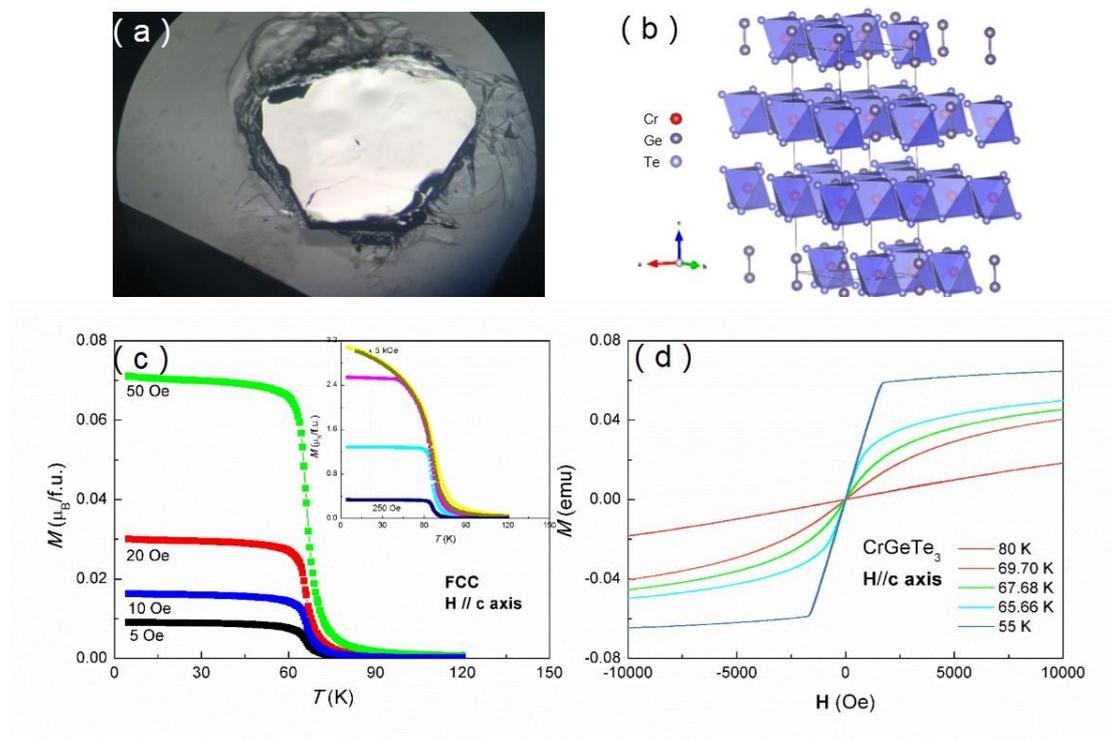

Fig.1 (a) $Cr_2Ge_2Te_6$ bulk sample glued on thin $SiO_2$ plate for MFM measurement. (b) shows the crystal structure of $Cr_2Ge_2Te_6$. (c, d) SQUID measurements of the *M-T* and *M-H* curves for the $Cr_2Ge_2Te_6$ sample. The Curie temperature is about 65 K.

The sample was glued on a $SiO_2$ substrate, and was exfoliated with Scotch tape to obtain new clean cleavage surface for MFM imaging. We used our homebuilt MFM to record the magnetic domain structure and evolution, which could be inserted into the cold bore of an 18/20 T superconducting magnet (from Oxford Instruments). The force sensors used were piezoresistive cantilevers (PRC400 from Hitach High-Tech Science Corporation, Japan) whose pyramid-shaped tips were coated with sandwiched magnetic thin films (5 nm Cr buffer layer followed by 50 nm Co and then 5 nm Au). They were then magnetized with a permanent magnet.

At first, the dynamic evolution of magnetic domain at microscale were obtained by taking MFM images in a temperature range covering the Curie point. Results are demonstrated in Fig. 2, from which some important features can be identified as follows:

(1) From 60 to 67.5 K, the dark regions overall form a pattern of large connected triangles (LCTs). This type of symmetry is hence called LCTs for convenience. Sometimes, a group of three LCTs share a common vertex and are equally spaced in a circle (called Y-connected LCTs), as indicated by the green triangles in Fig. 2a and Fig. 5d; and sometimes, LCTs are connected in series, forming a zigzag pattern (called Zigzag-LCTs), as marked by the green dashed line in Fig. 5e.

(2) There are numerous small disk-like domains of varying sizes embedded in the large dark or bright spacial regions. They may be hollow polygons if they are not too small in size, which will be shown more clearly later in Fig. 4d. These disks are embedded and distributed in a self-fitting (SF) manner: their sizes and distribution density are self-adjusted to well fit the local space and shape of the region they occupy. This can be more clearly seen in the image taken at 67.5 K. This type of symmetry is called SF disks (SFDs). This self-fitting property is somewhat similar to the property of a standing wave. It is thus probable that the SF disks stem from the standing waves of spin density. The SFDs, to the knowledge of the authors, have not been reported in 3D materials, that means they might be unique to 2D materials where the standing wave of spin density could be easier to form. The highest value of SFD density is 3.13um$^{-2}$, which is sufficiently high for storage applications.

(3) At 68.9 K, the evolution comes to a stage where the bright regions become narrow and form lots of Y-shaped connects (YSCs) as shown in Fig. 2d. The pivot (center) at which three walls are connected can be either a solid dot or hollow polygon (Fig.5b). The connected walls actually have fine structure of ladder type as shown by the pattern in the boxed area of Fig. 2d whose zoom-in scan is given in Fig.5b for a better view. The number of SFDs is now greatly reduced.

(4) As the temperature increases further, the YSCs are connected into a maze

pattern (MP), then fade away and disappear as the sample becomes paramagnetic. The SFDs become completely invisible prior to the disappearance of the MP. Actually, just before the MP disappears, it becomes less isotropic and more skewed (or aligned) toward the direction of the PPSs. This can be better seen when the temperature decreases from 69.3 K (where the MP has become fully disappeared) and reaches 69.1 and 68.9 K (Fig.2), respectively. It is worth emphasizing that the walls of the MP also have ladder type fine structures (Fig. 2l, Fig.2m) that are typically missing in 3D materials and may be unique to 2D materials. After the MP vanishes, there are still some periodic parallel stripes (PPSs) remaining visible. The latter will play an important role as we will see later on.

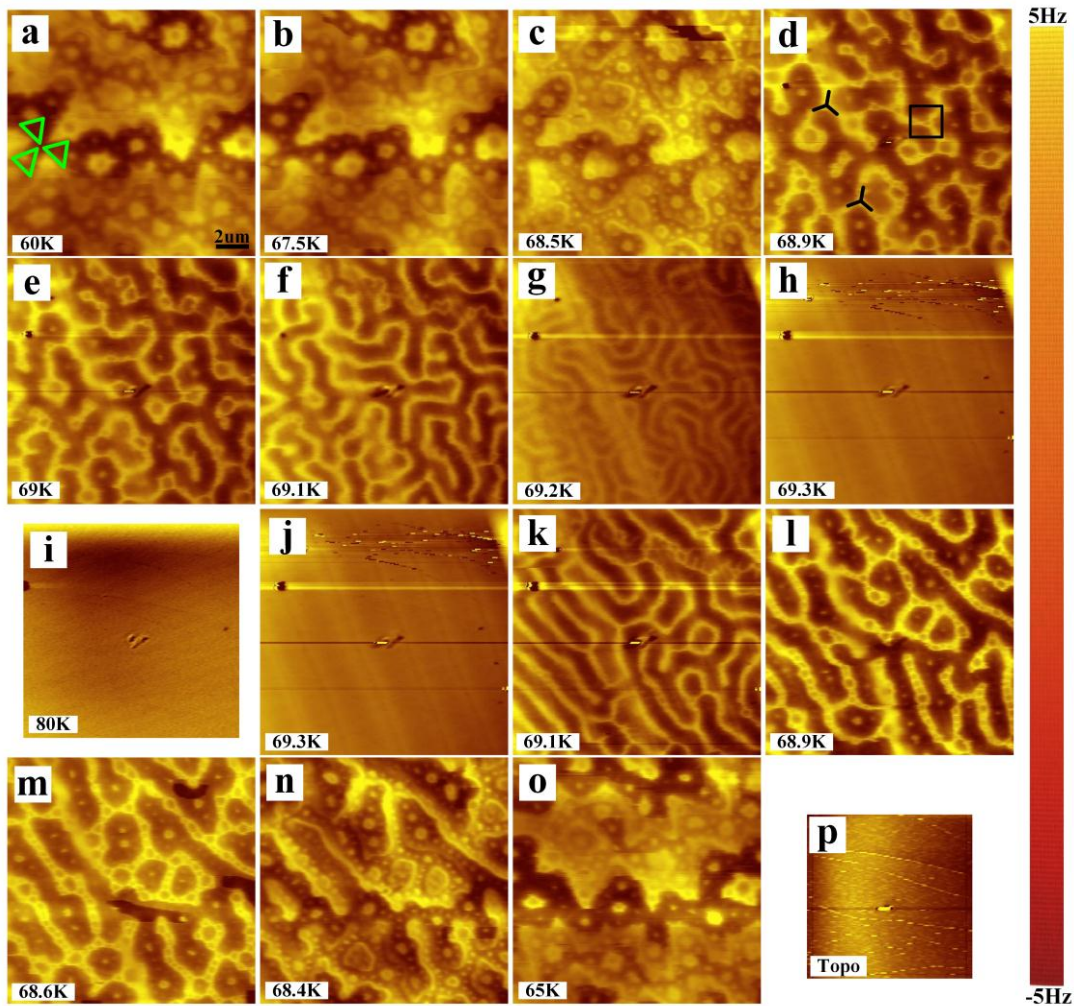

Fig.2. MFM images showing the temperature dependence of magnetic domain structures and symmetries at the surface of the bulk $Cr_2Ge_2Te_6$ crystal sample. Scan area of (a-h, j-o): 15μm × 15μm. Scan area of (i): 12μm × 12μm. (a) - (i) were taken in a temperature increasing process. A Y-connected LCT is illustrated with green triangles in (a). Y-shaped connects are shown with black "Y" in (d). (i) - (o) were obtained in a temperature decreasing process. The right-bottom image is the topography scanned in a reduced area (9 μm × 9 μm) suggesting quite high quality of the sample surface.

(5) When the temperature decreases, the symmetry type changes roughly in a reverse manner (with a small degree of hysteresis), namely, from PPSs (69.3 K), to MP without SFDs (69.1 K), to YSCs with a small number of SFDs (68.9 to 68.6 K) and to LCTs with a large number of SFDs (65 K). During this process, we are able to see the impact of the direction of the PPSs for most of the time.

In general, the overall pattern becomes less symmetric and more isotropic as the temperature increases. It is worth emphasizing that all these plentiful symmetry types are experienced within a small temperature range of 2 K or so. This tells that different symmetries are competing in a subtly balanced manner, which might be related to the drastic competition between thermal fluctuation and interlayer coupling.

We also imaged the dynamic evolution of domain structure driven by the applied magnetic field (Fig. 3) at 69.9 K. The first image (Fig. 3b) in the series shows YSC-type symmetry with a small number of SFDs. As the external magnetic field increases, the number of SFDs is reduced and the pattern gets more severely impacted by the direction of the PPSs.

Surprisingly, when the magnetic field increases to 0.04 T, the pattern becomes much of PPS type and a new pattern shows up, namely, the periodic dot lattice (PDL) type pattern with hexagonal symmetry which is more clearly seen under 0.05 T. Apparently, the PDL pattern is superimposed on the PPS pattern with the dots now appearing on the bright PPS lines. The density of the dots is about 0.14 $\mu m^{-2}$ which is in the applicable range for data storage.

When the field goes slightly further to 0.06 T, the bright dots become small dark rings. At 0.07 T, a step of just 0.01 T increase, the dots completely disappear, with only the pattern of PPSs left. Thus, it does not require strong magnetic field to manipulate these dots, which is definitely good news for storage applications.

The above phenomena (movie) can be played in a reverse order if we decrease the strength of the magnetic field without changing its direction. When the magnetic field is reduced from 0.07 T, small dark rings appear first, then bright dots at 0.04 T. But YSCs do not come into being with further reduction of magnetic field all the way to zero.

However, at -0.01 T (the direction of magnetic field is reversed) YSCs appear and even grow into a network. The resulting pattern looks like a honeycomb (HC) with equally spaced bright dots in each cell. Then the adjacent cells merge into larger ones but the dots are gone at -0.02 T. As the field gets stronger, the HC pattern loses contrast and the PPS pattern shows up. Comparing Fig. 3(j) and (k), we find that the pivots of the YSCs tend to located at the PPS lines to become the dots of PDL as the magnetic field gets more negative. Finally everything except the PPSs are killed by a high field of -0.07 T, just as expected.

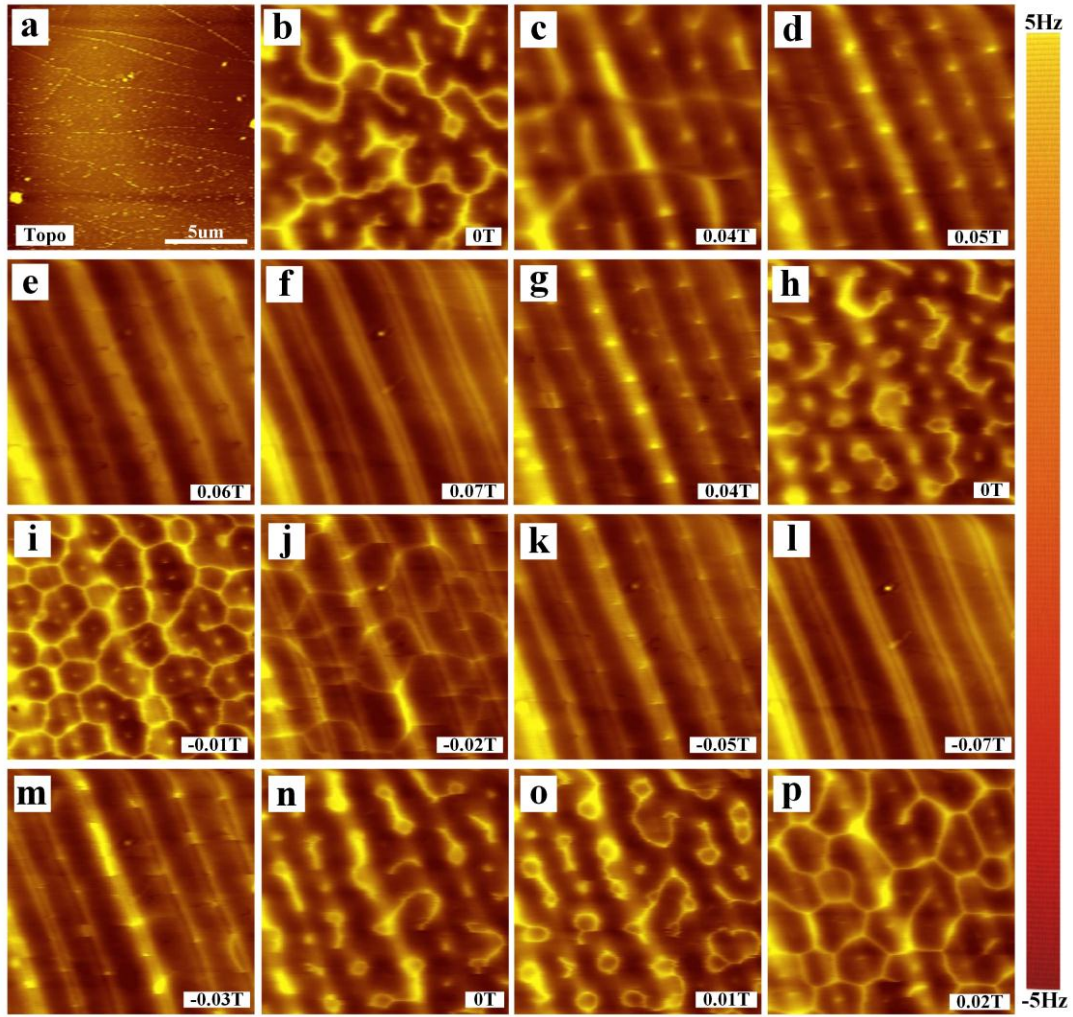

Fig.3. (a) The topographic image of the scanned area (15μm×15μm). (b) - (p) Evolution of magnetic structure and symmetry as a function of external magnetic field at 69.9K.

In general, the overall pattern becomes more symmetric and less isotropic as the magnetic field increases. It is also worth emphasizing that all these plentiful symmetry types are experienced within a small magnetic field range of 0.07 T (700 Gauss), which implies great chance for practical applications.

Now we have found that so many types of magnetic structures and symmetries exist or co-exist in the 2D magnetic crystal. Their basic structures are picked up and summarized in Fig. 4, including (a) stripe, (b) maze, (c) dot, (d) disk, (e) line-connected disk, (f) YSC, (g) HC, (h) SFD and (i) multiple-layered disk. Apparently, some of the domain pattern symmetries are correlated with the real-space symmetry of the crystal structure (Fig.1b, Fig.1c), in which the triangular shape,

Y-connection, zigzag pattern and hexagonal symmetry all exist. They can all show up in a narrow range of temperature and magnetic field, probably because of the close competition between thermal fluctuation and anisotropy energy induced by magnetic dipolar interaction.

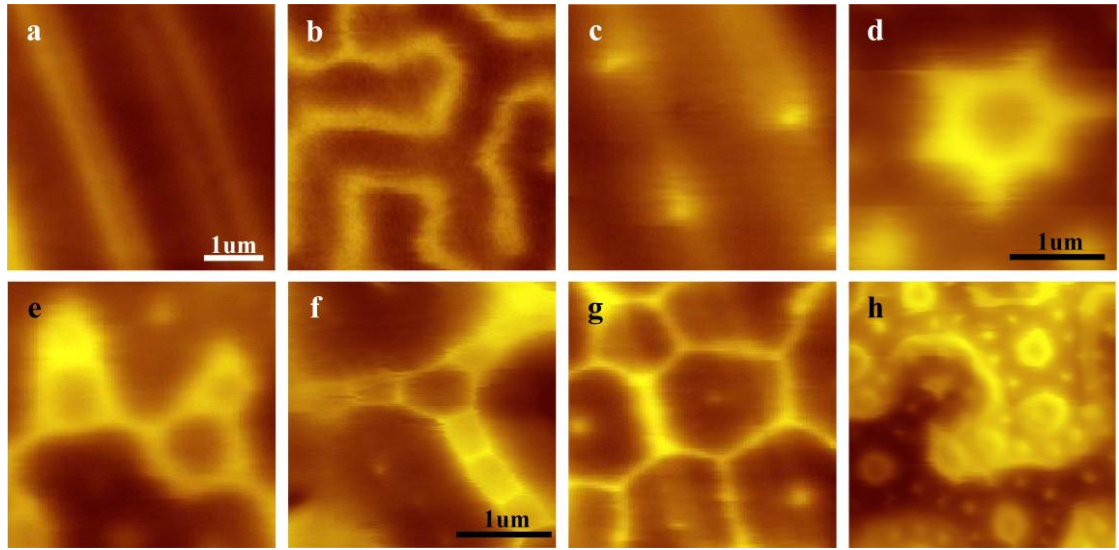

Fig. 4. MFM images for eight kinds of magnetic structures obtained in the $Cr_2Ge_2Te_6$ crystal. **a**) stripe, **b**) maze, **c**) dot, **d**) disk, **e**) line-connected disk, **f**) YSC, **g**) HC, **h**) SFD and multiple-layered disk. Image sizes are 5×5 μm$^2$ except d), f) whose image sizes are 3×3 μm$^2$.

The transitions between these structures or symmetries can be multiform and informative too. In addition to the aforementioned ones, some new or more detailed forms are listed in Fig. 5. Fig. 5a shows how dots in PDL are transformed into individual disks and then connected HC cells with a dot at the center. Fig. 5b illustrates how MP evolves into a series of YSCs in which a zoom-in scanned one has hollow polygon pivot and fine ladder structure in the walls. The most informative transformation is given in Fig. 5c, which demonstrates how the fine ladder structure in a certain domain wall can change into a series of homologous SFDs; Namely, each ladder cell transforms into a corresponding isolated SFD. Also, we can find that when a certain large SFD shrinks, the small ones nearby will grow large to occupy the extra space available. This well reflects the self-fitting nature of the SFDs. Fig. 5d and e show the aforementioned Y-connected LCT and zigzag-LCT, respectively. Fig. 5f shows the schematic spin configuration for the Y-connected LCT structure. Areas marked with "+" and "-" are spin-up and spin-down regions, respectively, so they show dark (attractive force) and bright (repulsive force) contrasts in the scanned image.

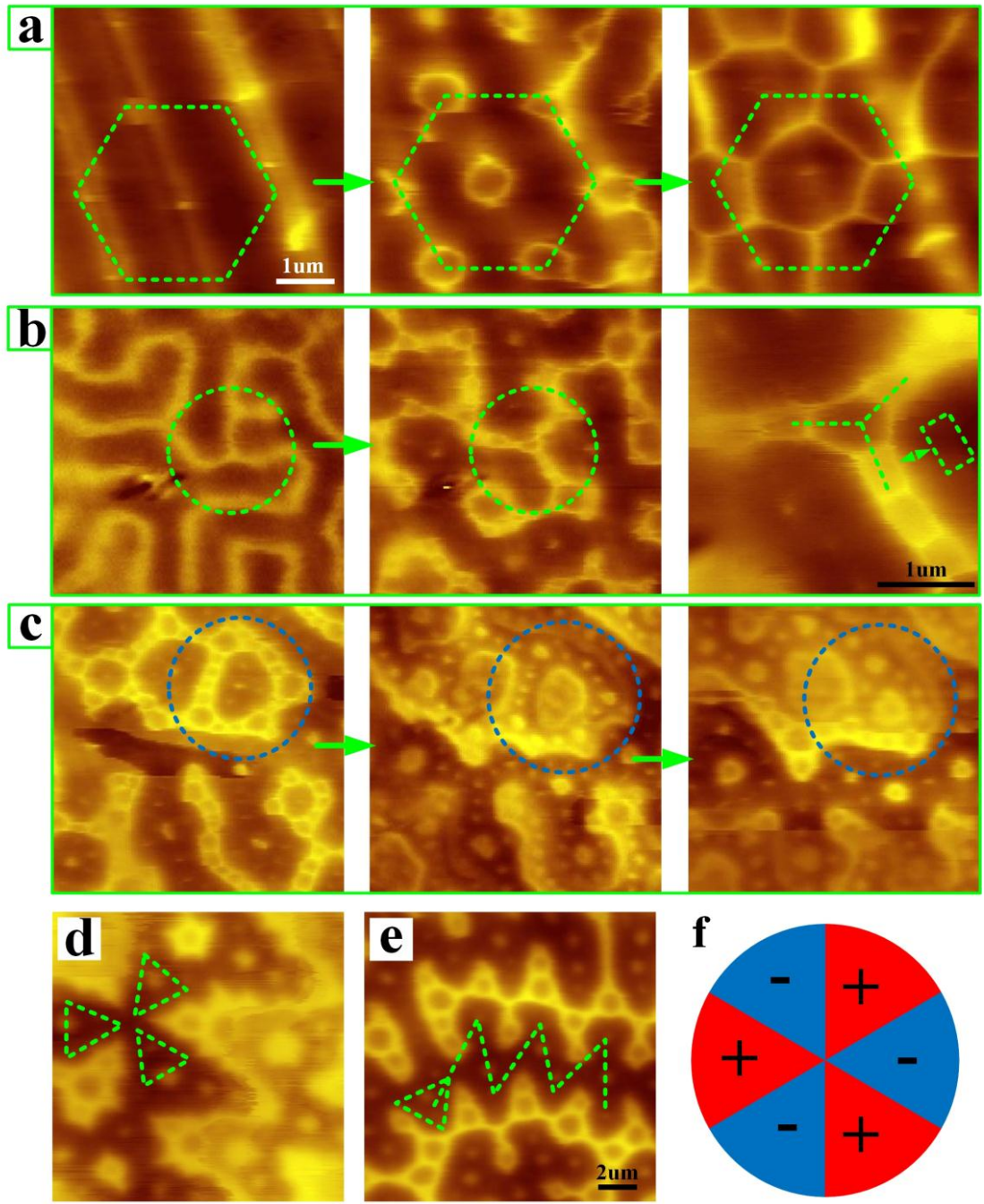

Fig. 5. a) MFM images showing magnetic phase evolution from dots to individual

circles, and then to fully touched circles. b) Formation and zoom-in imaging of fine ladder on Y-connected walls. c) Magnetic phase evolution of fine ladder structure into self-fitting disks. d) Y-connected LCT. e) Zigzag-connected triangles. (f) A cartoon explaining the possible ferromagnetic polarity (up or down) arrangement in a Y-connected LCT domain.

The above abundant types of magnetic structures and symmetries mutually compete or impact with each other, depending on how we control the temperature and applied magnetic field. Based on these important data, we can construct a structure or symmetry phase diagram for the ferromagnetic phase along, in which the X and Y axes are magnetic field and temperature, respectively. This is shown in Fig. 6. Basically, it reflects a global picture of high symmetry level at lower temperature and higher magnetic field. After all, lower temperature means less thermal disturbance, which is in favor of retaining the intrinsic symmetry of the lattice, hence showing high anisotropy too. Higher magnetic field is also prone to evoke anisotropy as it is a highly directional field.

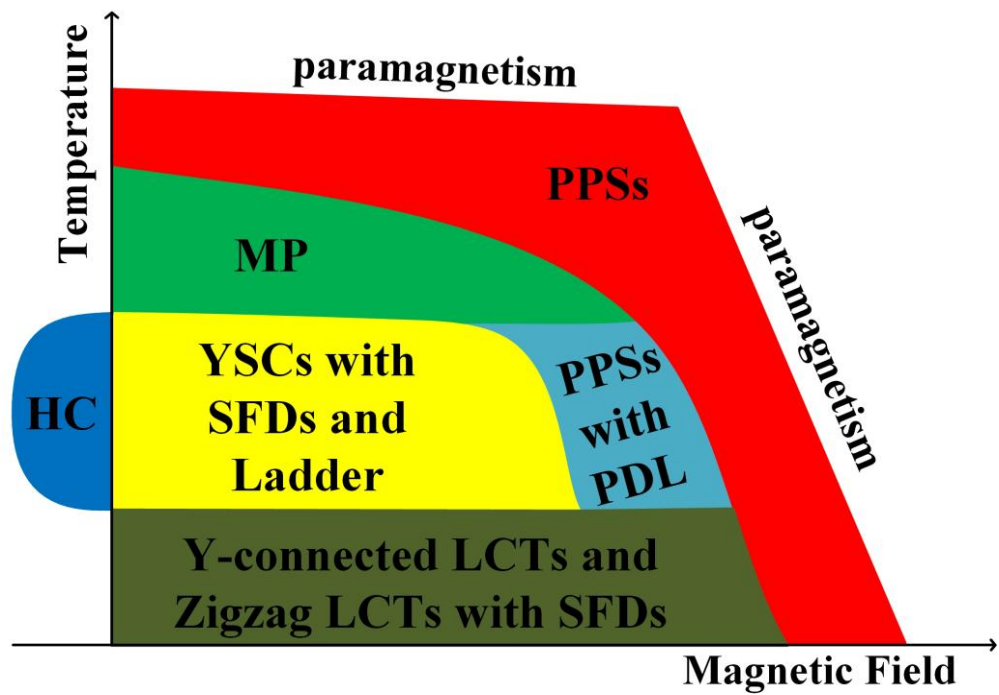

Fig.6. Schematic drawing of the structure (or symmetry) phase diagram of the ferromagnetic phase in the quasic-2D $Cr_2Ge_2Te_6$ sample.

## 3 Conclusion

In a two-dimensional magnetic material $Cr_2Ge_2Te_6$, we have successfully observed many magnetic domain structures, classified their symmetry types and revealed how they transform in responding to temperature and applied magnetic field. The domain structures are plenty and their evolution processes are very informative. They can mutually transform within a very narrow temperature range of only 2 K and a very narrow magnetic field range of only 0.07 T. All these picturesque characteristics constitute the unique long-range 2D magnetism behavior, which show high potential of practical applications as well as fundamental research value.

**Acknowledgement**

This work was supported by the National Key R&D Program of China (Nos. 2017YFA0402903 and 2016YFA0401003), the National Natural Science Foundation of China (Nos. 21505139, 51627901, 11474263, 11674326, U1432139 and U1432251), the Anhui Provincial Natural Science Foundation (No. 1608085MB36), the Dean fund of Hefei institutes of Physical Science of CAS (No.YZJJ201620), and the Chinese Academy of Sciences Scientific Research Equipment (No. YZ201628).